\documentclass[reprint,aip,floatfix]{revtex4-1}
\pdfoutput=1

\usepackage{amsmath}
\usepackage{paralist}

\usepackage{graphicx}
\graphicspath{ {./} {imgpdf/} {imgeps/} }
\usepackage{tikz}
\usetikzlibrary{arrows}
\usetikzlibrary{spy}
\usepackage{pgfplots}

\pgfplotsset{compat=1.5}
\pgfplotsset{
emphasize/.code args={#1:#2with#3}{
    \pgfplotsextra{
            \fill[#3] ({axis cs:#1,0} |- {axis description cs:0,0}) 
            rectangle ({axis cs:#2,0} |- {axis description cs:0,1});
        }
    }
}
\pgfplotsset{
	colormap={matlab}{
		gray=( 1.000000)
		gray=( 0.985714)
		gray=( 0.971429)
		gray=( 0.957143)
		gray=( 0.942857)
		gray=( 0.928571)
		gray=( 0.914286)
		gray=( 0.900000)
		gray=( 0.885714)
		gray=( 0.871429)
		gray=( 0.857143)
		gray=( 0.842857)
		gray=( 0.828571)
		gray=( 0.814286)
		gray=( 0.800000)
		gray=( 0.785714)
		gray=( 0.771429)
		gray=( 0.757143)
		gray=( 0.742857)
		gray=( 0.728571)
		gray=( 0.714286)
		gray=( 0.700000)
		gray=( 0.685714)
		gray=( 0.671429)
		gray=( 0.657143)
		gray=( 0.642857)
		gray=( 0.628571)
		gray=( 0.614286)
		gray=( 0.600000)
		gray=( 0.585714)
		gray=( 0.571429)
		gray=( 0.557143)
		gray=( 0.542857)
		gray=( 0.528571)
		gray=( 0.514286)
		gray=( 0.500000)
		gray=( 0.485714)
		gray=( 0.471429)
		gray=( 0.457143)
		gray=( 0.442857)
		gray=( 0.428571)
		gray=( 0.414286)
		gray=( 0.400000)
		gray=( 0.385714)
		gray=( 0.371429)
		gray=( 0.357143)
		gray=( 0.342857)
		gray=( 0.328571)
		gray=( 0.314286)
		gray=( 0.300000)
		gray=( 0.285714)
		gray=( 0.271429)
		gray=( 0.257143)
		gray=( 0.242857)
		gray=( 0.228571)
		gray=( 0.214286)
		gray=( 0.200000)
		gray=( 0.185714)
		gray=( 0.171429)
		gray=( 0.157143)
		gray=( 0.142857)
		gray=( 0.128571)
		gray=( 0.114286)
		gray=( 0.100000)
	}
}
\newlength{\columnwidthae}
\usepackage{bibunits}
\defaultbibliography{ae_references,mainNotes}
\defaultbibliographystyle{unsrt}

\newcommand{\suppmat}[1][]{%
  \footnotetext{
    See attached supplementary material for 
    details  of the computation of the critical Mason number 
    and its asymptotic approximations.}%
  \newcounter{suppmatcite}\setcounter{suppmatcite}{\value{footnote}}%
  \renewcommand{\suppmat}[1][]{\cite{Note\arabic{suppmatcite}%
    \if\relax\detokenize{##1}\relax\else,##1\fi}}%
  \suppmat[#1]%
}

\begin{document}
\title{Particle Capture Efficiency in a Multi-Wire Model for 
	High Gradient Magnetic Separation} 
\author{Almut Eisentr\"ager}
\author{Dominic Vella}
\author{Ian M. Griffiths}
\email[]{ian.griffiths@maths.ox.ac.uk}
\affiliation{Mathematical Institute,
University of Oxford,
Radcliffe Observatory Quarter,
Woodstock Road,
Oxford,
OX2 6GG,
United Kingdom}
\date{\today}
\begin{abstract}
High gradient magnetic separation (HGMS) is an efficient way to 
remove magnetic and paramagnetic particles, 
such as heavy metals, from waste water. 
As the suspension flows through a magnetized filter mesh, 
high magnetic gradients around the wires attract and capture the particles, removing them from the fluid. 
We model such a system by considering the motion of
a paramagnetic tracer particle through a periodic array of magnetized cylinders. 
We show that there is a critical \emph{Mason number}
(ratio of viscous to magnetic forces) 
below which the particle is captured irrespective of its initial position in the array. 
Above this threshold, particle capture is only partially successful 
and depends on the particle's entry position. 
We determine the relationship between the critical Mason number and the system geometry using numerical and 
asymptotic calculations. 
If  a capture efficiency below 100\% is sufficient, 
our results demonstrate how
operating the HGMS system
above the critical Mason number but with multiple separation cycles
may increase efficiency.
\end{abstract}
\maketitle

\begin{bibunit}
Various applications require efficient removal of magnetic and paramagnetic particles from a carrier fluid{,
such as waste water treatment, food processing and microfluidics}.{\cite{yavuz2009,ambashta2010,hayashi2011a,sinha2007,tsai2011a}}
In high gradient magnetic separation (HGMS),  {a} suspension  {flows} through a  filter
made of magnetized material, 
such as  regular mesh grids or randomly packed material (steel wool) 
in the field of a strong electromagnet.
\cite{metso-spec-cyclic}
The particles are deflected by magnetic forces due to the strong magnetic field 
gradients between the filter wires which enables particle capture within the filter.
Despite these techniques remaining
effectively unchanged since the 1970s,
\cite{pfister1979,gerber1983}
several theoretical questions  remain unanswered.

The wire volume fraction in a typical HGMS system (2--15\%) is well below what would be necessary for mechanical filtration. 
Nevertheless, the magnetic and hydrodynamic interactions between wires can play an important role in the trajectory of contaminant particles 
and whether they are captured. 
However, many previous attempts to model HGMS systems focus on the ability of a single wire 
to capture a single particle\cite{cummings1976,gerber1983,ebner2001,mishima2012} 
or to retain large numbers in the late stages of filtration.\cite{uchiyama1977a,chenf2012,lindner2013b}
To model the effects of many wires, single-wire results are often superposed\cite{too1986,kimyg2013}
or particular geometries and parameter values are studied.\cite{hayashi1980,simons1980,mariani2010,hayashi2011a}
These studies reveal that for potential flow within a periodic square lattice of cylinders, particles may escape filtration 
if they enter in a narrow escape trajectory whose width depends on the geometry, strength of magnetic interactions and viscous drag. 
In this Letter we focus on providing a complete understanding of this dependence, focusing in particular on the role of the packing density of the wires, 
which has not been systematically considered before, and how filtration efficiency can be maximized.

\begin{figure}
{
\begin{tikzpicture}[thick,x=0.65cm,y=0.65cm,>=latex]
    \foreach \r in {0.6}
    {
      \foreach \x in {-3,-1,...,3}
      {
	\foreach \y in {-2,0,...,2}
	{
	  \draw[fill=lightgray] (\x,\y) circle (\r);
	  \draw[->] (\x-0.3,\y) -- (\x+0.3,\y);
	}
      }

      \node at (-1,2) [above] {$M$};

      \draw[thin,dashed,->] (0,-2.75) -- (0,3);
      \node at (0,3) [above] {$y$};

      \draw[thin,dashed,->] (-4,0) -- (4,0);
      \node at (4,0) [right] {$x$};

      \draw [|-|] (3.2+\r,2-\r) -- (3.2+\r,2+\r);
      \node at (3.2+\r,2) [right] {$2R$};
      
      \draw [|-|] (3.2+\r,-\r) -- (3.2+\r,-2+\r);
      \node at (3.2+\r,-1) [right] {$2d$};

      \draw [|-|] (1,2.25+\r) -- (3,2.25+\r);
      \node at (2,2.25+\r) [above] {$2L=2(R+d)$};

      \draw [thick,dotted] (0,-1) -- (1,-1) -- (1,1) -- (0,1) -- (0,-1);

      \draw[->,thin] (0.3,-1)  to[out=120,in=190]  (0.4,0);
      \draw[->,thin] (0.3,-1)  to[out=120,in=-100]  (0.2,0.4);
      \fill (0.3,-1) circle (0.15);
      \draw[lightgray,->] (0.3-0.15,-1) -- (0.3+0.15,-1);
      \node at (0.3,-1.1) [below] {$m$};

      \draw[->, ultra thick] (0,0) -- (0,0.8);
      \node at (0.05,0.7) [left] {$\!U$};
      
      \draw[->, ultra thick] (-2,-3) -- (2,-3);
      \node at (0,-3) [below] {strong uniform magnetic field};
    }

  \end{tikzpicture}
}  \caption{
    Cylinder array and particle 
    in Setup A (top view).
    The dotted box indicates the computational domain.
  }
  \label{fig:setup}
\end{figure}
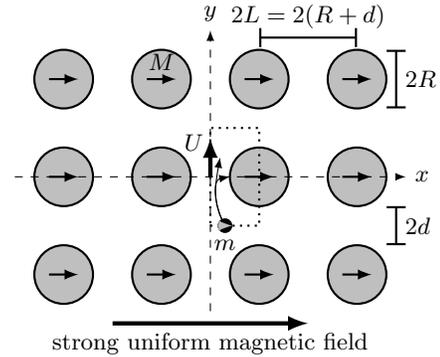
As a simplified model for the filter material, 
here we  consider a large square array of long parallel cylinders
of radius $R$
with a constant magnetization perpendicular to their axes
(Fig.~\ref{fig:setup}). 
The magnetic dipole moment of the particle is constant and kept aligned with the magnetization
of the cylinders  {by} the {action of a} strong, uniform outer magnetic field.
The smallest distance between cylinders in the $x$- and $y$-direction
is denoted by $2d$, so that the period of the array in both directions
is ${2L=2(R+d)}$.
 {Using} symmetry across the $y$-axis,
we  {may} reduce the computational domain to half 
the periodic cell,
  (Fig.~\ref{fig:setup}).
{We assume that the filtrate is}
a dilute suspension of magnetic particles. 
This allows us to
neglect interactions between particles and
focus on a single spherical magnetic particle 
moving through this cylinder array, carried by a 
fluid flow and deflected by the magnetic force exerted 
{by} the magnetic field of the cylinders. 
{This model captures the important physics of the system and represents a worst case scenario,
since higher particle concentrations would cause chains to form that would be captured more easily.\cite{yavuz2006}}

We consider two different setups. 
In Setup~A, the magnetization of the cylinders and the particle are 
perpendicular to the flow direction (Fig.~\ref{fig:setup}). 
In Setup~B, both magnetizations are  parallel to the flow direction.
For a given setup and operating conditions, we are interested
in whether a particle entering the computational domain at the  {inlet} with
$(x_\text{p}(0),y_\text{p}(0))=(x_0,-L)$ 
can escape, that is, leave the domain at some
$(x_\text{p}(t),y_\text{p}(t))=(x_\text{p}(t),L)$,
or whether 
{its trajectory intersects the cylinder,
in which case we say it has been captured}.

For the calculations, we make the following assumptions.
\begin{inparaenum}[(a)]
\item The cylinders are infinitely long and the flow field is planar, \emph{i.e.},
    two-dimensional.
\item The cylinder array is infinite, thus both the flow field and the magnetic field
    can be considered  {to be} periodic.
 \item The flow is steady and laminar.
 \item The particle diameter, $2a$, is small 
  compared to
  the smallest distance between cylinders, $2d$,
  that is, $a/d\ll1$,
  and so the particle does not disturb the flow field.
\end{inparaenum}

In practical applications, the operating Reynolds number ranges from 
$10^{-4}$ (\emph{e.g.},  in food processing)  to several hundreds 
(\emph{e.g.}, in waste water treatment).
\cite{hayashi2011a,kimyg2013}
Here we are concerned with
full particle capture, which can be achieved 
either by increasing the strength of the magnetization 
or by decreasing the flow rate.
Since the latter is more feasible in most cases,
we assume we are in the lower Reynolds number range 
and model the fluid flow as Stokes flow. 
However, the following analysis {might} readily  {be extended to}
potential flow or the full Navier--Stokes equations.

Neglecting inertial terms, 
the net force on the particle must be zero, 
\emph{i.e.}, the magnetic force, $\boldsymbol{F}_\text{m}$,
and the viscous drag force (Stokes drag)
must balance, giving
\begin{align}
 6\pi\eta a \left[ \boldsymbol{u}(\boldsymbol{x}_\text{p})-\dot{\boldsymbol{x}}_\text{p}\right]+\boldsymbol{F}_\text{m}&=0,
\label{eq:dimforcebalance}
\end{align}
with $\eta$ the  fluid viscosity and   $\boldsymbol{x}_\text{p}$ the particle position.
The dot $\dot{~}$ denotes differentiation with respect to time.
The force exerted on a particle due
to a single magnetic cylinder is given by\suppmat
\begin{align}
 \boldsymbol{F}_\text{m, single}&=\mp\frac{\mu_0 mM}{R}\left(\frac{R}{r}\right)^3
  \begin{pmatrix}
   \cos3\theta\\\sin3\theta
  \end{pmatrix},
\end{align}
{where $\mp$ corresponds to} Setup~A/B, respectively,
$\mu_0$ {is} the permeability of free space, 
$M$ the magnetization of the cylinders, 
$m$ the magnetic dipole moment of the particle,  
and $r$ and $\theta$ the plane-polar coordinates centered at the cylinder midpoint.
The total magnetic force on a particle is the sum of the contributions from
all cylinders in the array.

Upon nondimensionalizing the system,
we find that the behavior of the particle is governed  only
by its initial position and the so-called Mason number,
which measures the strength of the viscous forces  compared 
with the magnetic forces in the system, 
\suppmat[kang2012]
\begin{align}
 \text{Mn}&
  =\pm\frac{6\pi\eta a RU}{\mu_0 mM}
\label{eq:MasonNumber}
\end{align}
for Setup~A and B, respectively.
Here the velocity scale, $U$, is taken as the maximum fluid velocity,
occurring at the origin, \emph{i.e.}, at  the midpoint between neighboring cylinders.

\begin{figure}
    \begin{tikzpicture}
    \begin{axis}
      [
      ,small
      ,enlargelimits=false
      ,axis on top
      ,scale mode=scale uniformly
      ,axis equal image
      ,width=0.7\columnwidthae
      ,every x tick/.style={black}
      ,tick align=center
      ,xtick={0,0.333,1},xticklabels={0 , $x_\text{c}$ , $L$}
      ,ytick={-1,0,1},yticklabels={$-L$,0,$L$}
      ,point meta min=0, point meta max=1.1
      ,mark options={scale=2}
      ]
    \addplot graphics
    [xmin=0,xmax=1,ymin=-1,ymax=1]
    {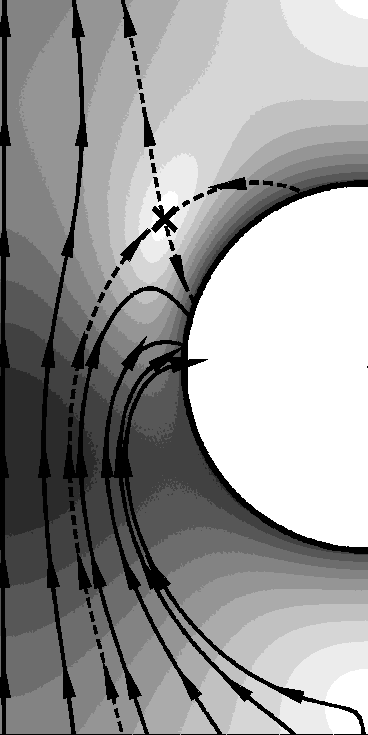};
    
    \addplot[ultra thick,mark=x] coordinates {(0.4474, 0.4024)};

    \end{axis}
    \draw[thick, ->] (1.2,1.84) -- (1.6,1.84);
    \node at (1.4,1.84) [above] {$M$};
    \draw [thick,->] (-0.6,1.84-0.5) -- (-0.6,1.84+0.5);
    \node at (-0.6,1.84) [left] {$U$};

    \begin{axis}
      [
      ,small
      ,enlargelimits=false
      ,axis on top
      ,scale mode=scale uniformly
      ,axis equal image
      ,width=0.7\columnwidthae
      ,every x tick/.style={black}
      ,tick align=center
      ,xtick={0,0.333,1},xticklabels={0 , $x_\text{c}$ , $L$}
      ,ytick={-1,0,1},yticklabels={$-L$,0,$L$}
      ,yticklabels=\empty
      ,colorbar
      ,colorbar style={ytick={0,0.5,1},yticklabels={0,$U/2$,$U$}}
      ,point meta min=0, point meta max=1.1
      ,mark options={scale=2}
      ,xshift=2.5cm
      ]
    \addplot graphics
    [xmin=0,xmax=1,ymin=-1,ymax=1]
    {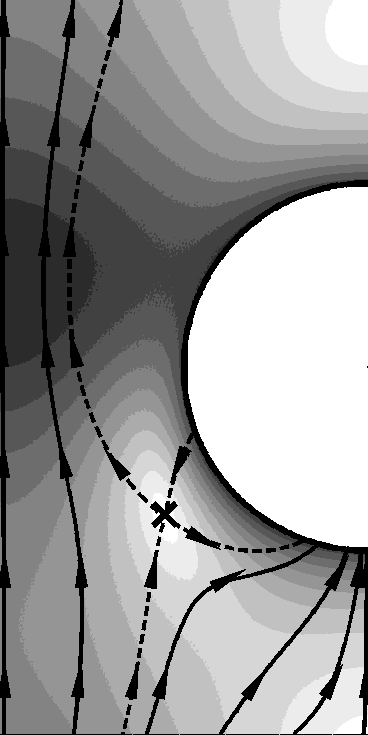};
    \addplot[ultra thick,mark=x] coordinates {( 0.4473 ,  -0.4025)};
    \end{axis}
    \draw [thick, ->] (1.6+2.5,1.84-0.2) -- (1.6+2.5,1.84+0.2);
    \node at (1.6+2.5,1.84) [left] {$M$};

    \node at (1.55,3.4)     {\textbf{(a)}};
    \node at (1.55+2.5,3.4) {\textbf{(b)}};

    \end{tikzpicture}
  \caption{
    Example particle trajectories (black curves)
    through a single periodic cell of the cylinder array
    and 
    dimensionless particle speed (gray shading) as a function of position
    for dimensionless cylinder radius $d/R=1$ and
    ${\text{Mn}=1}$:
    (a)~Setup~A,
    (b)~Setup~B.
    The cross  ($\boldsymbol{\times}$)
    indicates the critical point between escape and capture. 
    This point is 
    {a stationary point that is unstable to lateral perturbations}.
    The dashed trajectories indicate the critical trajectories
    starting and ending in the critical point.
    The value $x_\text{c}$ denotes the critical initial $x$-position
    between capture and escape at the  {inlet to}
    the periodic cell.
  }
  \label{fig:trajectories}
\end{figure}
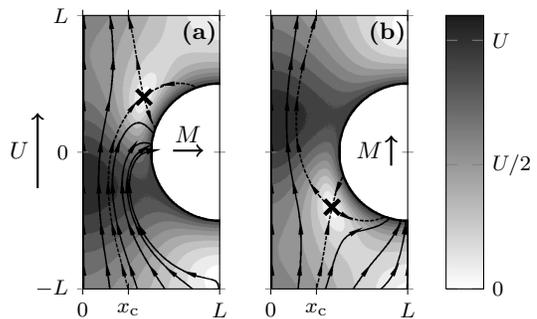
We solve the flow problem of periodic Stokes flow
through an infinite, regular cylinder array
numerically with the Finite Element Method in FreeFEM++
for  {a range of}  cylinder radii {and spacings}.
\cite{freefemppdocumentation}
Using this fluid velocity and approximating the magnetic field of an infinite cylinder array
by that of a sufficiently large finite array (in practice a $10\times19$ array is sufficient),
 we can compute 
the particle velocity for any particle position
and thus numerically integrate the particle trajectory 
from any given initial position
(Fig.~\ref{fig:trajectories}). 
In the case of a moderate cylinder magnetization, 
the particle can escape if it starts
close enough to the  midline between
two cylinders.
Along these trajectories, the magnetic force on the particle
is too weak to overcome the viscous drag force.
Since both
the fluid flow and the magnetic force
are periodic in $y$,
these trajectories have to be periodic
and so  a particle that escapes one periodic cell 
of the cylinder array will also escape all subsequent cells.
\cite{simons1980}

If a particle enters the cell closer to the cylinder,
the stronger magnetic force and lower drag force 
due to slower fluid speeds closer to the cylinder wall
both result in a trajectory that is more strongly influenced by the magnetic field.
In Setup~A,
the particle is repelled from the front of the cylinder
and attracted to its side where it is eventually captured
(Fig.~\ref{fig:trajectories}(a)).
In Setup~B,
the particle is instead attracted to,
 and captured at, the front of the cylinder 
(Fig.~\ref{fig:trajectories}(b)).

At the critical point, where the trajectories diverge between escape and capture,
a particle would have zero speed, as indicated by the white background 
in Fig.~\ref{fig:trajectories}. 
{This stationary point is unstable with respect to lateral displacements.}
Comparing the two setups, the position of this critical point 
is mirrored along the 
$x$-axis.
As the absolute value of the Mason number is decreased, 
the critical point moves from the cylinder wall to the $y$-axis.

\begin{figure}
\begin{tikzpicture}
\begin{axis}
  [
  ,small
  ,enlargelimits=false
  ,width=0.8\columnwidth
  ,height=0.5\columnwidth
  ,ylabel={efficiency / \%}
  ,xlabel={$|\text{Mn}|$}
  ,xmode=log,ymode=normal
  ,xmin=1e-3,xmax=1e3
  ,ymin=0,ymax=1.04,
  ,ymin=0,ymax=104,
  ]

  \addplot [thick,
    dash pattern=on 3pt off 2pt 
    ] coordinates { 
 (10, 0.85563)   (7.9433, 0.9919)   (6.3096, 1.213)   (5.0119, 1.49)  
 (3.9811, 1.7588)   (3.1623, 2.1047)   (2.5119, 2.5663)   (1.9953, 3.0138)   (1.5849, 3.6373)  
 (1.2589, 4.2682)   (1, 5.007)   (0.79433, 5.8772)   (0.63096, 6.8505)   (0.50119, 7.9515)  
 (0.39811, 9.1873)   (0.31623, 10.5909)   (0.25119, 12.1539)   (0.19953, 13.9157)   (0.15849, 15.8664)  
 (0.12589, 18.051)   (0.1, 20.4257)   (0.079433, 23.0341)   (0.063096, 25.9009)   (0.050119, 29.0034)  
 (0.039811, 32.369)   (0.031623, 35.999)   (0.025119, 39.8793)   (0.019953, 44.0262)   (0.015849, 48.4394)  
 (0.012589, 53.0945)   (0.01, 58.0038)   (0.0079433, 63.1398)   (0.0063096, 68.4942)   (0.0050119, 74.0099)  
 (0.0039811, 79.6555)   (0.0031623, 85.3164)   (0.0031623, 85.3154)   (0.0029854, 86.7115)   (0.0028184, 88.0958)  
 (0.0026607, 89.4618)   (0.0025119, 90.8048)   (0.0025119, 90.8023)   (0.0023714, 92.1189)   (0.0022387, 93.3943)  
 (0.0021135, 94.6278)   (0.0019953, 95.8001)   (0.0019953, 95.7978)   (0.0018836, 96.8965)   (0.0017783, 97.9017)  
 (0.0016788, 98.7772)   (0.0015849, 99.4861)   (0.0015849, 99.4862)   (0.0014962, 99.9035)   (0.0014125, 99.9591)  
 (0.0013335, 99.9607)   (0.0012589, 99.9622)   (0.0012589, 99.9626)   (0.0011885, 99.9641)   (0.001122, 99.9645)  
 (0.0010593, 99.9646)   (0.001, 99.9651)   (0.001, 99.9653)  }; 

\addplot [thick,solid] coordinates { 
 (100, 3.9601)   (79.4328, 4.8317)   (63.0957, 6.0497)   (50.1187, 7.3708)  
 (39.8107, 9.0839)   (31.6228, 11.0845)   (25.1189, 13.3546)   (19.9526, 15.8883)   (15.8489, 18.6587)  
 (12.5893, 21.7347)   (10, 24.9574)   (7.9433, 28.3338)   (6.3096, 31.959)   (5.0119, 35.731)  
 (3.9811, 39.6404)   (3.1623, 43.7795)   (2.5119, 47.9866)   (1.9953, 52.3799)   (1.5849, 56.9996)  
 (1.2589, 61.74)   (1, 66.7019)   (0.79433, 71.7885)   (0.63096, 77.0456)   (0.50119, 82.4218)  
 (0.39811, 87.7875)   (0.31623, 92.982)   (0.31623, 92.982)   (0.29854, 94.2101)   (0.28184, 95.3878)  
 (0.26607, 96.5074)   (0.25119, 97.5483)   (0.25119, 97.5483)   (0.23714, 98.469)   (0.22387, 99.2571)  
 (0.21135, 99.8213)   (0.19953, 99.9457)   (0.19953, 99.9457)   (0.18836, 99.9517)   (0.17783, 99.953)  
 (0.16788, 99.9554)   (0.15849, 99.9577)   (0.15849, 99.9577)   (0.14962, 99.9595)   (0.14125, 99.9601)  
 (0.13335, 99.9619)   (0.12589, 99.9624)   (0.12589, 99.9624)   (0.11885, 99.9626)   (0.1122, 99.9626)  
 (0.10593, 99.9632)   (0.1, 99.9645)   (0.1, 99.9645)  };

  \addplot [thick,
    ,dashdotted] coordinates { 
 (1000, 46.1503)   (794.3282, 47.334)   (630.9573, 47.6526)   (501.1872, 48.1937)  
 (398.1072, 49.346)   (316.2278, 50.2223)   (251.1886, 51.3651)   (199.5262, 52.4135)   (158.4893, 53.7196)  
 (125.8925, 55.1925)   (100, 55.8366)   (79.4328, 58.5111)   (63.0957, 60.0984)   (50.1187, 61.9664)  
 (39.8107, 64.0468)   (31.6228, 66.178)   (25.1189, 68.4743)   (19.9526, 70.9042)   (15.8489, 73.5339)  
 (12.5893, 76.2639)   (10, 79.2033)   (10, 79.3268)   (9.4406, 79.9973)   (8.9125, 80.7999)  
 (8.414, 81.5562)   (7.9433, 82.3726)   (7.9433, 82.3804)   (7.4989, 83.2143)   (7.0795, 84.0329)  
 (6.6834, 84.8406)   (6.3096, 85.7136)   (6.3096, 85.7229)   (5.9566, 86.5516)   (5.6234, 87.427)  
 (5.3088, 88.2935)   (5.0119, 89.179)   (5.0119, 89.1791)   (4.7315, 90.0861)   (4.4668, 91.011)  
 (4.217, 91.8673)   (3.9811, 92.7831)   (3.9811, 92.7756)   (3.7584, 93.678)   (3.5481, 94.5829)  
 (3.3497, 95.446)   (3.1623, 96.2888)   (3.1623, 96.2973)   (2.9854, 97.1232)   (2.8184, 97.9215)  
 (2.6607, 98.6431)   (2.5119, 99.2357)   (2.5119, 99.2357)   (2.3714, 99.747)   (2.2387, 99.931)  
 (2.1135, 99.94)   (1.9953, 99.9432)   (1.9953, 100)   (1.8836, 99.9476)   (1.7783, 99.9491)  
 (1.6788, 99.9523)   (1.5849, 99.9556)   (1.5849, 99.9552)   (1.4962, 99.956)   (1.4125, 99.9559)  
 (1.3335, 99.9569)   (1.2589, 99.9588)   (1.2589, 99.959)   (1.1885, 99.9603)   (1.122, 99.9604)  
 (1.0593, 99.961)   (1, 99.9618)   (1, 99.9617)  }; 

  \end{axis}

  \end{tikzpicture}
  \caption{
    Influence of the Mason number on 
    the capture efficiency, $1-x_\text{c}/L$,
    for different geometries:
    $d/R=9$ (dashed), $d/R=1$ (solid), $d/R=1/9$ (dashdotted).
  }
\label{fig:percentagecapturedperMn}
\end{figure}
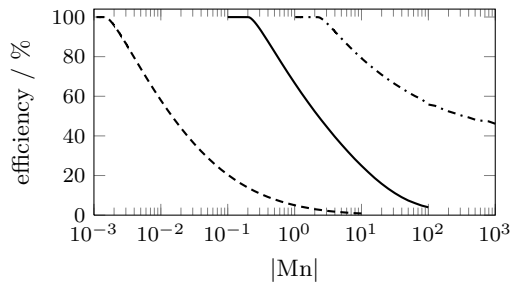

We denote by  $x_\text{c}$ the critical initial $x$-position,
\emph{i.e.}, the initial $x$-position of the critical trajectory that leads from 
the  {inlet to} the cell to the critical point.
Particles with initial position $x<x_\text{c}$
will escape, those with $x\ge x_\text{c}$ will get captured.\footnote{
  Due to our choice of coordinate system, 
  the ``capture distance'' or ``capture cross section'' 
  from the literature\cite{cummings1976,simons1980}
  is $L-x_\text{c}$.
}
Surprisingly, the value of $x_\text{c}$
only depends on the absolute value of the Mason number,
despite the very dissimilar limiting trajectories 
for positive and negative Mason numbers
(Fig.~\ref{fig:trajectories}).
{Assuming the filtrate is well mixed},
the capture efficiency  is  given by $100(1-x_\text{c}/L){\%}$
and only depends on the absolute value of the Mason number and the geometry
(Fig.~\ref{fig:percentagecapturedperMn}).

For each geometry, 
there exists 
a critical absolute Mason number,   $\text{Mn}_\text{crit}$,
below which no particle can escape, 
regardless of its initial position at the  {inlet to} the cell,
and all particles are instead captured at the side or front of the cylinder 
depending on the setup (Fig.~\ref{fig:percentagecapturedperMn}).
The particle whose initial position lies on the midline between the two cylinders,
\emph{i.e.}, $(x_\text{p}(0),y_\text{p}(0))=(0,-L)$, 
is the particle that is most easily able to escape,
and will thus determine the critical Mason number.
Due to symmetry, both the magnetic force and the fluid velocity have only 
 components
in the $y$-direction along this line
and a particle that originates at $(0,-L)$ will,  in theory,
remain on the $y$-axis for all time. 
Thus, to find the critical Mason number, 
we may restrict our focus to the one-dimensional problem of whether or not a 
particle travelling along  the $y$-axis escapes.
In practice, 
instabilities or diffusion might move the particle away from the $y$-axis,
so that it is captured even at higher Mason numbers,
but the one-dimensional problem considered here provides an upper bound for the
critical Mason number.

For Mason numbers with absolute value below the critical Mason number,
the critical point lies on the $y$-axis
and the particle velocity is negative along parts of this axis. 
For values above the critical Mason number, 
the particle velocity needs to be positive along the whole $y$-axis
for the particle to be able to escape.
Thus the critical Mason number is that absolute Mason number
for which the particle velocity on the $y$-axis just reaches the value zero
at some position.\suppmat{}
The capture efficiency and thus the critical Mason number 
depend on  the geometry, 
namely the ratio $d/R$ of 
the smallest distance between cylinders to
the cylinder diameter
(Fig.~\ref{fig:percentagecapturedperMn}),
which is  related to
the filter volume fraction, $\phi$, 
by $d/R=\sqrt{\pi/(4\phi)}-1$.

If the distances between cylinders are very small compared to their radii, 
that is $d/R\ll1$,
the flow field in the gap between two cylinders
can be approximated by lubrication theory.~\cite{ockendon1995}
In addition, 
since all other cylinders are further away and
thus contribute less to the overall magnetic force,
we consider only the influence 
of these two cylinders on the magnetic field as a first approximation.
Hence, we obtain 
the following asymptotic approximation for the 
critical Mason number \suppmat
\begin{align}
\begin{split}
 \text{Mn}_{\text{crit}}&=\frac{1}{216}\left(34\sqrt{2}+5\sqrt{5}\right)\frac{R}{d}
  \approx0.27\left(\frac{d}{R}\right)^{-1}
\end{split}
\end{align}
as $d/R\rightarrow 0$.

For distant cylinders, such that $d/R\gg1$,
the velocity  away from the cylinders along the $y$-axis
is approximately constant,
while the magnetic force along the $y$-axis
decreases as 
\begin{align}
   F_{\text{m},y}(0,y)&\sim 
  \left(\frac{\sqrt{(d+R)^2+y^2}}{R}\right)^{-3}
    \sim \mathcal{O}\left(\frac{d}{R}\right)^{-3}
\end{align}
as $d/R\rightarrow\infty$.
Thus, 
we obtain
 ${\text{Mn}_{\text{crit}}\sim \mathcal{O}\left(d/R\right)^{-3}}$
in this limit.\suppmat

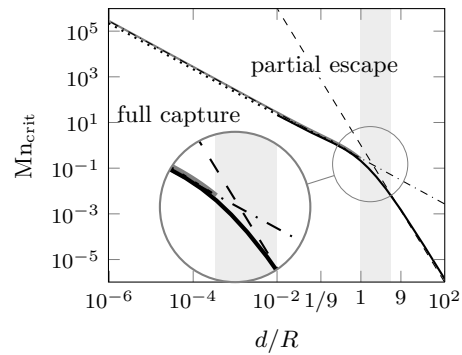
\begin{figure}
\begin{tikzpicture}[spy using outlines={circle,lens={scale=2}, connect spies}]
 \begin{loglogaxis}[
    legend style={cells={anchor=west},
    legend pos=outer north east},
    xlabel={$d/R$},ylabel={$\text{Mn}_\text{crit}$},
    xtick=      {0.01,      0.111111,  1, 9, 100,       1e4,      1e6},
    xticklabels={$10^{2}$, {9},     1, {1/9}, $10^{-2}$,  $10^{-4}$, $10^{-6}$},
    enlargelimits=false,
    small,
    axis on top,
    xmin=0.01   ,
    x dir=reverse,
    width=0.7\columnwidth,
  ]

\addlegendentry{Lubrication, 2 cylinders};
\addplot [gray,thick,emphasize=0.1899:1.0187 with black!07, 
  ] coordinates { 
 (1048576, 287696.7866)   (912838.4275, 250454.6693)   (794672.0072, 218033.5231)   (691802.1635, 189809.276)  
 (602248.7631, 165238.6418)   (524288, 143848.6623)   (456419.2137, 125227.6037)   (397336.0036, 109017.0306)   (345901.0818, 94904.907)  
 (301124.3816, 82619.5899)   (262144, 71924.6002)   (228209.6069, 62614.0708)   (198668.0018, 54508.7843)   (172950.5409, 47452.7225)  
 (150562.1908, 41310.064)   (131072, 35962.5691)   (114104.8034, 31307.3044)   (99334.0009, 27254.6612)   (86475.2704, 23726.6303)  
 (75281.0954, 20655.301)   (65536, 17981.5536)   (57052.4017, 15653.9212)   (49667.0005, 13627.5996)   (43237.6352, 11863.5841)  
 (37640.5477, 10327.9195)   (32768, 8991.0458)   (28526.2009, 7827.2296)   (24833.5002, 6814.0688)   (21618.8176, 5932.0611)  
 (18820.2738, 5164.2287)   (16384, 4495.7919)   (14263.1004, 3913.8838)   (12416.7501, 3407.3033)   (10809.4088, 2966.2995)  
 (9410.1369, 2582.3833)   (8192, 2248.1649)   (7131.5502, 1957.2108)   (6208.3751, 1703.9206)   (5404.7044, 1483.4186)  
 (4705.0685, 1291.4605)   (4096, 1124.3513)   (3565.7751, 978.8742)   (3104.1875, 852.229)   (2702.3522, 741.978)  
 (2352.5342, 645.9989)   (2048, 562.4443)   (1782.8876, 489.7057)   (1552.0938, 426.383)   (1351.1761, 371.2574)  
 (1176.2671, 323.2678)   (1024, 281.4904)   (891.4438, 245.121)   (776.0469, 213.4595)   (675.5881, 185.8966)  
 (588.1336, 161.9016)   (512, 141.0126)   (445.7219, 122.8277)   (388.0234, 106.9967)   (337.794, 93.2149)  
 (294.0668, 81.2171)   (256, 70.7722)   (222.8609, 61.6793)   (194.0117, 53.7632)   (168.897, 46.8717)  
 (147.0334, 40.8721)   (128, 35.6489)   (111.4305, 31.1014)   (97.0059, 27.1424)   (84.4485, 23.6954)  
 (73.5167, 20.6941)   (64, 18.0809)   (55.7152, 15.8053)   (48.5029, 13.8236)   (42.2243, 12.0976)  
 (36.7583, 10.5941)   (32, 9.2841)   (27.8576, 8.1425)   (24.2515, 7.1473)   (21.1121, 6.2793)  
 (18.3792, 5.5218)   (16, 4.8602)   (13.9288, 4.2819)   (12.1257, 3.7756)   (10.5561, 3.3318)  
 (9.1896, 2.9419)   (8, 2.5986)   (6.9644, 2.2954)   (6.0629, 2.0267)   (5.278, 1.7878)  
 (4.5948, 1.5744)   (4, 1.3832)   (3.4822, 1.2112)   (3.0314, 1.056)   (2.639, 0.91595)  
 (2.2974, 0.78951)   (2, 0.67562)   (1.7411, 0.57348)   (1.5157, 0.48247)   (1.3195, 0.40204)  
 (1.1487, 0.33168)   (1, 0.27081)  }; 

\addlegendentry{Lubrication, array};
\addplot [black,thick,dotted] coordinates { 
 (1048576, 224822.8313)   (262144, 56206.2367)   (65536, 14052.0881)   (16384, 3513.5508)  
 (4096, 878.9163)   (1024, 220.2566)   (256, 55.5875)   (64, 14.4036)   (16, 4.0418)  
 (4, 1.2326)   (1, 0.24742)  }; 

\addlegendentry{FEM solution, array};
\addplot [black,thick] coordinates { 
 (0.010101, 1.44e-06)   (0.030928, 3.94e-05)   (0.11111, 0.00149)   (0.25, 0.0122)  
 (0.42857, 0.0421)   (0.66667, 0.102)   (1, 0.207)   (1.5, 0.3755)   (2.3333, 0.64565)  
 (4, 1.12)   (9, 2.28)   (15.6667, 3.7)   (99, 21.2)  }; 

\addlegendentry{$\left(\frac{R}{d}\right)^3$};
\addplot [dashed] coordinates { 
 (0.010101, 1.0306e-06)   (0.030928, 2.9583e-05)   (0.11111, 0.0013717)   (0.25, 0.015625)  
 (0.42857, 0.078717)   (0.66667, 0.2963)   (1, 1)   (1.5, 3.375)   (2.3333, 12.7037)  
 (4, 64)   (9, 729)   (15.6667, 3845.2963)   (99, 970299)  }; 

\addlegendentry{$\alpha\frac{R}{d}$};
\addplot [dashdotted] coordinates { 
(0.010101, 0.0027273)   (0.030928, 0.0083505)   (0.11111, 0.03)   (0.25, 0.0675)  
 (0.42857, 0.11571)   (0.66667, 0.18)   (1, 0.27)   (1.5, 0.405)   (2.3333, 0.63)  
 (4, 1.08)   (9, 2.43)   (15.6667, 4.23)   (99, 26.73) 
(294.0668, 79.398)
(2352.5342, 635.1842) 
(18820.2738, 5081.4739)  
(150562.1908, 40651.7915) 
(602248.7631, 162607.166)  
(1048576, 283115.52)   
}; 

  \node at (10,3) {full capture};
  \fill[white] (2.5,8) circle[radius=1.2];
  \node at (2,8)  {partial escape};
  \legend{}

  \coordinate (spypoint) at (axis cs:0.6,0.15);
  \coordinate (magnifyglass) at (axis cs:1000,0.002);

 \end{loglogaxis}

\spy [gray, size=2cm] on (spypoint) in node[fill=white] at (magnifyglass);

\end{tikzpicture}
  \caption{Critical Mason number for different ratios of cylinder radius and spacing: 
    numerical computation using the finite element flow field and the  magnetic force
      from a cylinder array {(thick solid)},
    numerical computation using the lubrication flow field and the  magnetic force
      from a cylinder array {(thick dotted)},
    numerical computation using the lubrication flow field and the  magnetic force
      from only two cylinders {(thick gray)},
    asymptotic limit for small distances using the lubrication flow field and the magnetic force
      from only two cylinders {{(dash-dotted)}},
    asymptotic behavior for large distances {(dashed)}.
    The shading indicates the parameter range representative 
    of typical HGMS systems.\cite{cummings1976,kimyg2013}
    }
  \label{fig:Mncritical-all}
\end{figure}
For cylinder separations that are of the same order of magnitude 
as the cylinder radius, that is ${d/R=\mathcal{O}(1)}$,
numerical solutions of the fluid flow field must be used (here  
obtained with the Finite Element package FreeFEM++).\cite{freefemppdocumentation}
The critical Mason numbers obtained via numerical solutions are in excellent 
agreement with the asymptotic limits (Fig.~\ref{fig:Mncritical-all}).
If the setup is chosen such that the absolute value of the Mason number
is below the critical Mason number
then capture of all particles sent through the system can be guaranteed.
To lower the Mason number to achieve this, one can, for example,
 reduce the fluid  velocity
through the cylinder array. 
Alternatively, our results imply that closer packing 
of the filter material improves capture efficiency by 
increasing the critical Mason number (Fig.~\ref{fig:Mncritical-all}).
However, these improvements need to be weighed against the drop in flow rate
and the concomitant reduced rate of production of clean water
that this implies
or the necessary  increase in the pressure gradient to keep the flow rate the same.

\begin{figure}
  \begin{tikzpicture}
    \begin{axis}
      [
      ,small
      ,enlargelimits=false
      ,scale mode=scale uniformly
      ,scale=0.9
      ,ylabel={\#cycles}
      ,xlabel={$|\text{Mn}|$}
      ,xmode=log
      ,ymode=log
      ,ymax=400,ymin=0.8
      ,width=0.5\columnwidthae
      ]


    \addplot [thick,dashed] coordinates { 
    (10, 349)   (7.9433, 301)   (6.3096, 246)   (5.0119, 200)  
    (3.9811, 169)   (3.1623, 141)   (2.5119, 116)   (1.9953, 98)   (1.5849, 81)  
    (1.2589, 69)   (1, 59)   (0.79433, 50)   (0.63096, 43)   (0.50119, 37)  
    (0.39811, 32)   (0.31623, 27)   (0.25119, 24)   (0.19953, 20)   (0.15849, 18)  
    (0.12589, 16)   (0.1, 14)   (0.079433, 12)   (0.063096, 10)   (0.050119, 9)  
    (0.039811, 8)   (0.031623, 7)   (0.025119, 6)   (0.019953, 6)   (0.015849, 5)  
    (0.012589, 4)   (0.01, 4)   (0.0079433, 4)   (0.0063096, 3)   (0.0050119, 3)  
    (0.0039811, 2)   (0.0031623, 2)   (0.0025119, 2)   (0.0019953, 1)   (0.0015849, 1)  
    (0.0012589, 1)   (0.001, 1)  }; 

    \addplot [thick,solid] coordinates { 
    (100, 75)   (79.4328, 61)   (63.0957, 49)   (50.1187, 40)  
    (39.8107, 32)   (31.6228, 26)   (25.1189, 21)   (19.9526, 18)   (15.8489, 15)  
    (12.5893, 13)   (10, 11)   (7.9433, 9)   (6.3096, 8)   (5.0119, 7)  
    (3.9811, 6)   (3.1623, 6)   (2.5119, 5)   (1.9953, 5)   (1.5849, 4)  
    (1.2589, 4)   (1, 3)   (0.79433, 3)   (0.63096, 3)   (0.50119, 2)  
    (0.39811, 2)   (0.31623, 2)   (0.25119, 1)   (0.19953, 1)   (0.15849, 1)  
    (0.12589, 1)   (0.1, 1)  (0.01, 1)  }; 

    \addplot [thick,dashdotted] coordinates { 
 (1000, 5)   (794.3282, 5)   (630.9573, 5)   (501.1872, 5)  
 (398.1072, 5)   (316.2278, 5)   (251.1886, 5)   (199.5262, 5)   (158.4893, 4)  
 (125.8925, 4)   (100, 4)   (79.4328, 4)   (63.0957, 4)   (50.1187, 4)  
 (39.8107, 3)   (31.6228, 3)   (25.1189, 3)   (19.9526, 3)   (15.8489, 3)  
 (12.5893, 3)   (10, 2)   (7.9433, 2)   (6.3096, 2)   (5.0119, 2)  
 (3.9811, 2)   (3.1623, 1)   (2.5119, 1)   (1.9953, 1)   (1.5849, 1)  
 (1.2589, 1)   (1, 1)  (0.5, 1) }; 


    \end{axis}

    \begin{axis}
      [
      ,small
      ,enlargelimits=false
      ,scale mode=scale uniformly
      ,scale=0.9
      ,ylabel={$\text{\#cycles}/|\text{Mn}|$}
      ,xlabel={$|\text{Mn}|$}
      ,xmode=log
      ,ymode=log
      ,thick
      ,width=0.5\columnwidthae
      ,xshift=4.5cm
      ]

\addlegendentry{};
\addplot [dashed] coordinates { 
 (10, 34.9)   (7.9433, 37.8937)   (6.3096, 38.9884)   (5.0119, 39.9052)  
 (3.9811, 42.4509)   (3.1623, 44.5881)   (2.5119, 46.1804)   (1.9953, 49.1163)   (1.5849, 51.1075)  
 (1.2589, 54.8086)   (1, 59)   (0.79433, 62.9463)   (0.63096, 68.1504)   (0.50119, 73.8247)  
 (0.39811, 80.3804)   (0.31623, 85.3815)   (0.25119, 95.5457)   (0.19953, 100.2374)   (0.15849, 113.5723)  
 (0.12589, 127.0925)   (0.1, 140)   (0.079433, 151.071)   (0.063096, 158.4893)   (0.050119, 179.5736)  
 (0.039811, 200.9509)   (0.031623, 221.3594)   (0.025119, 238.8643)   (0.019953, 300.7123)   (0.015849, 315.4787)  
 (0.012589, 317.7313)   (0.01, 400)   (0.0079433, 503.5702)   (0.0063096, 475.468)   (0.0050119, 598.5787)  
 (0.0039811, 502.3773)   (0.0031623, 632.4555)   (0.0025119, 796.2143)   (0.0019953, 501.1872)   (0.0015849, 630.9573)  
 (0.0012589, 794.3282)   (0.001, 1000)  }; 


\addlegendentry{};
\addplot [solid] coordinates { 
 (100, 0.75)   (79.4328, 0.76794)   (63.0957, 0.7766)   (50.1187, 0.7981)  
 (39.8107, 0.8038)   (31.6228, 0.82219)   (25.1189, 0.83603)   (19.9526, 0.90214)   (15.8489, 0.94644)  
 (12.5893, 1.0326)   (10, 1.1)   (7.9433, 1.133)   (6.3096, 1.2679)   (5.0119, 1.3967)  
 (3.9811, 1.5071)   (3.1623, 1.8974)   (2.5119, 1.9905)   (1.9953, 2.5059)   (1.5849, 2.5238)  
 (1.2589, 3.1773)   (1, 3)   (0.79433, 3.7768)   (0.63096, 4.7547)   (0.50119, 3.9905)  
 (0.39811, 5.0238)   (0.31623, 6.3246)   (0.25119, 3.9811)   (0.19953, 5.0119)   (0.15849, 6.3096)  
 (0.12589, 7.9433)   (0.1, 10)   (0.01, 100) }; 


\addlegendentry{};
\addplot [dashdotted] coordinates { 
 (1000, 0.005)   (794.3282, 0.0062946)   (630.9573, 0.0079245)   (501.1872, 0.0099763)  
 (398.1072, 0.012559)   (316.2278, 0.015811)   (251.1886, 0.019905)   (199.5262, 0.025059)   (158.4893, 0.025238)  
 (125.8925, 0.031773)   (100, 0.04)   (79.4328, 0.050357)   (63.0957, 0.063396)   (50.1187, 0.07981)  
 (39.8107, 0.075357)   (31.6228, 0.094868)   (25.1189, 0.11943)   (19.9526, 0.15036)   (15.8489, 0.18929)  
 (12.5893, 0.2383)   (10, 0.2)   (7.9433, 0.25179)   (6.3096, 0.31698)   (5.0119, 0.39905)  
 (3.9811, 0.50238)   (3.1623, 0.31623)   (2.5119, 0.39811)   (1.9953, 0.50119)   (1.5849, 0.63096)  
 (1.2589, 0.79433)   (1, 1)   (0.5, 2) }; 


    \legend{}
    \end{axis}

  \node at (2.1    ,1.7) {\textbf{(a)}};
  \node at (2.1+4.5,1.7) {\textbf{(b)}};

  \end{tikzpicture} 

  \caption{
    {Conditions} on
    (a)~number of cycles,
    (b)~separation time, $t_\text{sep}\propto\text{\#cycles}/|\text{Mn}|$,
    to ensure $95\%$ capture efficiency
    for different Mason numbers and different geometries:
     ${d/R=9}$ (dashed), ${d/R=1}$ (solid), ${d/R=1/9}$ (dashdotted).
  }
  \label{fig:remove95percent}
\end{figure}
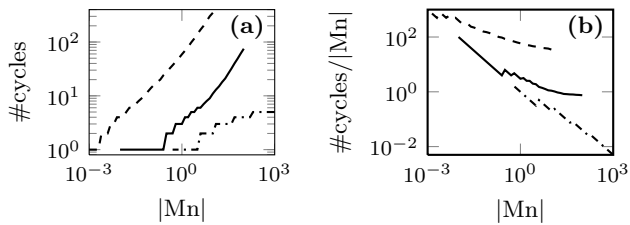

If full capture is required, the system must operate below the the 
critical Mason number. 
If, however, the required capture efficiency is lower, say only 95\%,
then for any given geometry and Mason number, it is possible
to achieve this by repeating the separation several times,
which may be faster 
than doing a single cycle at a lower Mason number.
Before each separation cycle we assume that the suspension is mixed again 
to randomize the initial $x$-position of the particles.
The separation cycles are then independent and we can infer how 
many cycles are necessary from the capture efficiency of
a single separation cycle 
(Fig.~\ref{fig:remove95percent}(a)).

Since the absolute value of the Mason number
depends linearly on the flow velocity, \eqref{eq:MasonNumber},
doubling the flow rate through the system doubles the Mason number.
Hence this increases the number of cycles necessary 
to achieve a required capture efficiency,
but also halves the time each separation cycle lasts.
Thus by dividing the number of necessary cycles 
by the absolute value of the Mason number, 
we can infer the effect of changing the Mason number
on the total separation time, $t_\text{sep}\propto\text{\#cycles}/|\text{Mn}|$,
neglecting any additional time that might be necessary between cycles
(Fig.~\ref{fig:remove95percent}(b)).
It is clearly inefficient to run the system below 
the highest Mason number that achieves the required capture efficiency or 
just above a Mason number at which the required number of cycles increases by one.
Furthermore, our data suggests that it would be overall more efficient to 
increase the flow rate and adapt the number of cycles accordingly.
However, there are some caveats that must be considered.
Firstly, a higher flow rate requires higher pressure gradients, which in turn
implies a higher energy consumption.
Secondly, we have not included any time for the tasks that may be necessary 
between separation cycles,
such as re-mixing the solution.
Lastly, in many industrial HGMS systems,
the filter material is randomly packed rather than periodic.
In this case, unlike in our setups, 
particles that escape the first wires in a filter 
may well be captured further downstream.
Thus, the increasing number of cycles necessary to achieve 
a required capture efficiency at higher Mason numbers
would then simply translate to increasing the length of the filter.

We have demonstrated two potential methods that guarantee a certain 
required particle capture efficiency in 
a periodic model of high gradient magnetic separation.
We have computed the critical Mason number,
below which full particle capture can be ensured,
numerically and asymptotically,
and shown how this depends on the filter geometrical parameter $d/R$.
If the required capture efficiency is below 100\%, 
then this can be achieved by repeating the separation process several times {at $\text{Mn}>\text{Mn}_\text{crit}$}.
We have shown how the number of necessary cycles depends on the Mason number 
and the geometry and that it may be overall more efficient to 
carry out multiple separation cycles at a higher flow rate.
The results of this work should be useful in advancing  strategies for the removal
of magnetic or paramagnetic particles.

\vspace{0.5em}
This publication was based on work supported in part by Award No KUK-C1-013-04, 
made by King Abdullah University of Science and Technology (KAUST) 
and by Award No 113/277 made by the John Fell Fund.

\end{bibunit}

\begin{bibunit}
\onecolumngrid\vspace{2cm}
{\noindent \hfill\Large  Supplementary material\hfill}
\newcommand{\unitvector}[1]{\ensuremath{\boldsymbol{e}_{#1}}}
\newcommand{\Mn}{\ensuremath{\operatorname{Mn}}}
\newcommand{\Mnc}{\ensuremath{\operatorname{Mn}_\text{crit}}}
\renewcommand{\L}{\ensuremath{\sqrt{Rd}}}

\section{Problem description}
We consider a single spherical magnetic particle 
moving through 
an infinite square array of infinitely long parallel cylinders
of radius $R$
with a constant magnetization perpendicular to their axes.
The particle is carried by a 
fluid flow and deflected by the magnetic force exerted 
due to the magnetic field of the cylinders
(Fig.~\ref{fig:setup}).
We study two different setups. 
In Setup~A, the magnetization of the cylinders and the particle are 
perpendicular to the flow direction (Fig.~\ref{fig:setup}). 
In Setup~B, both magnetizations are  parallel to the flow direction.
The flow is assumed to be planar, steady, periodic Stokes flow through the array. 

\section{Magnetic force on particle around a single cylinder}
We approximate the magnetic force on the particle in the inifinite cylinder array 
by adding the contributions of single cylinders in a sufficiently large
finite array. First consider setup~A.

\begin{figure}\centering
    \centering
    \begin{tikzpicture}
    \begin{axis}%
      [,small
      ,enlargelimits=false
      ,axis on top
      ,scale mode=scale uniformly
      ,axis equal image] 
       \addplot graphics
	[xmin=-5,xmax=5,ymin=-5,ymax=5]
	{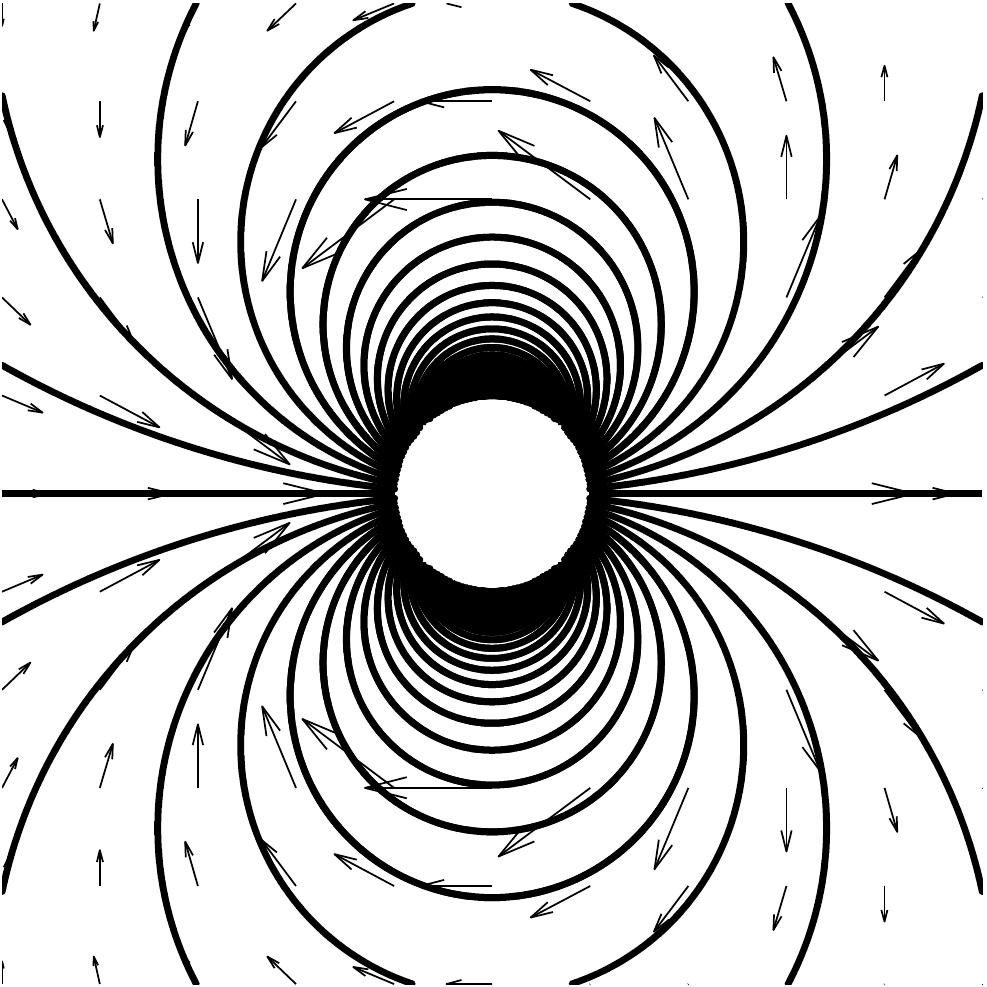};
    \end{axis}

    \begin{axis}%
      [,small
      ,enlargelimits=false
      ,axis on top
      ,scale mode=scale uniformly
      ,xshift=6cm
      ,axis equal image] 
       \addplot graphics
	[xmin=-5,xmax=5,ymin=-5,ymax=5]
	{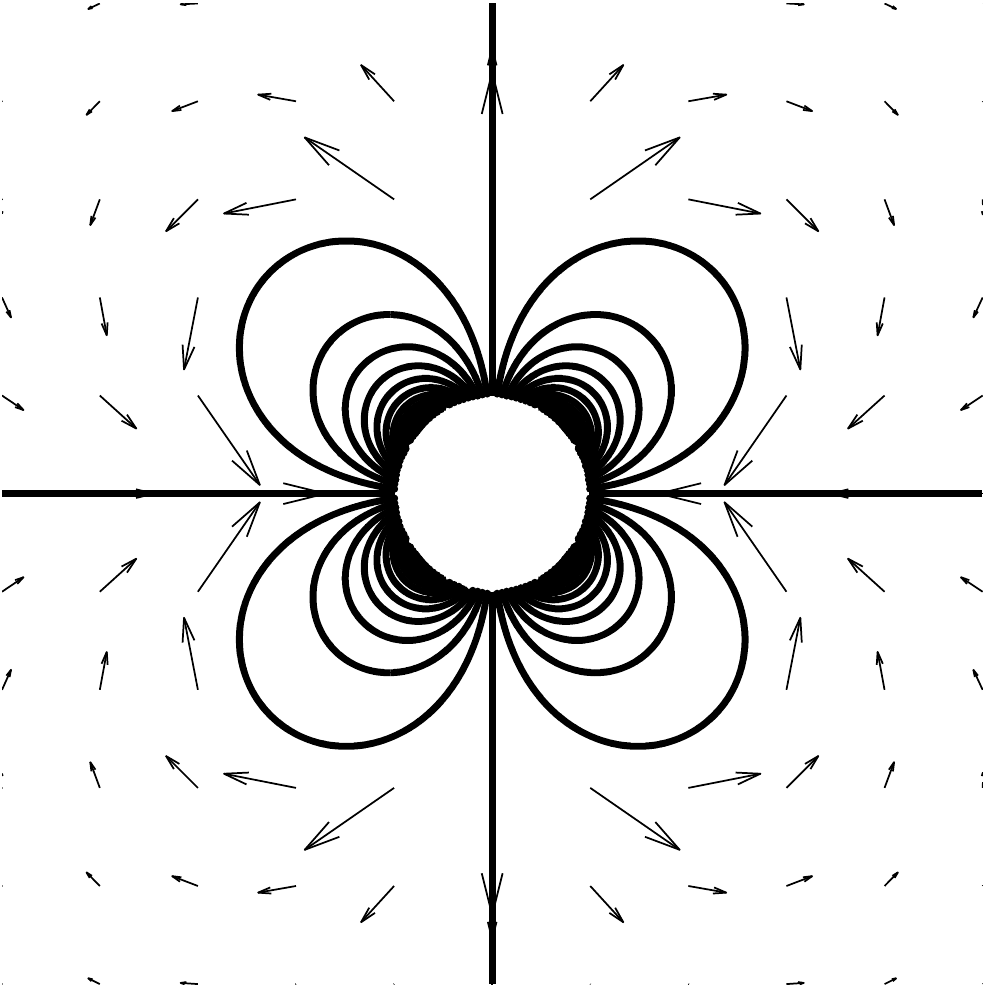};
    \end{axis}
    \node at (3.8  ,3.8)  {\textbf{(a)}};
    \node at (3.8+6,3.8)  {\textbf{(b)}};
    \end{tikzpicture}
  \caption{
    (a)~The magnetic field, $\boldsymbol{B}$, around a cylinder with uniform magnetization, 
      ${\boldsymbol{M}=M\unitvector{x}}$,
      perpendicular to its axis.
    (b)~The magnetic force, $\boldsymbol{F}$, on a particle around this cylinder
      with its magnetic dipole moment, $\boldsymbol{m}=m\unitvector{x}$, aligned
      to the cylinder magnetization.
  }
\label{supp-fig:magnfieldforce_singlecylinder}
\end{figure}

\begin{figure}
\centering
\begin{tikzpicture}%
  [thick,scale=0.7,
  >=latex]
  
  \newcommand{\R}{1.5}
  \newcommand{\x}{4}
  \newcommand{\y}{3}
  \newcommand{\s}{5} 
  \renewcommand{\u}{1.5} 

  \draw[fill=lightgray] (0,0) circle (\R);
  \draw [thick] (0,0) -- (0,-\R);
  \node at (0,-\R/2) [left] {$R$};
  
  \draw[fill=darkgray] (\x,\y) circle (0.15);
  
  \draw [thick] (0,0) -- (\x,0) -- (\x,\y) -- (0,0);
  \node at (\x/2,0) [below] {$x$};
  \node at (\x,1.5) [right] {$y$};
  \node at (\x/2,1.5) [above ] {$r$};
  \node at (\R/2,\R/4*\y/\x)  {$\theta$};

  \draw [ultra thick,->] (\x,\y) -- (\x +\u,\y);	 
  \node at (\x +\u,\y) [right] {\unitvector{x}};
  \draw [ultra thick,->] (\x,\y) -- (\x,\y +\u);	 
  \node at (\x,\y +\u) [above] {\unitvector{y}};
  \draw [ultra thick,->] (\x,\y) -- (\x + \u*\x/\s,\y + \u*\y/\s);	 
  \node at  (\x + \u*\x/\s*0.9,\y + \u*\y/\s*0.8) [above right] {\unitvector{r}};
  \draw [ultra thick,->] (\x,\y) -- (\x - \u*\y/\s,\y + \u*\x/\s);	 
  \node at  (\x - \u*\y/\s*0.8,\y + \u*\x/\s*0.9) [above left] {\unitvector{\theta}};

  \draw [dashed,->] (-\R -0.5,0) --(\x+\u+0.5,0);
  \node at (\x+\u+0.5,0) [right] {$x$};
  \draw [dashed,->] (0,-\R -0.5) --(0,\y+\u+0.5);
  \node at (0,\y+\u+0.5) [above] {$y$};
\end{tikzpicture}
\caption{Coordinate systems for single cylinder.}
\label{supp-fig:coordinates-cylinder}
\end{figure}
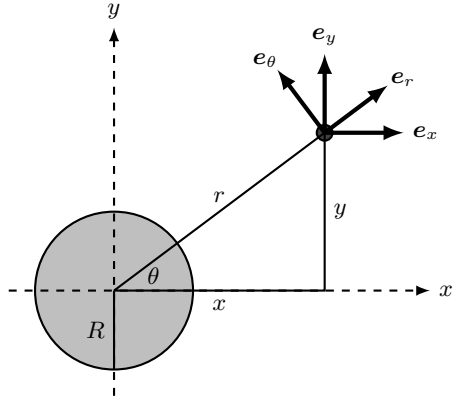

The magnetic field, $\boldsymbol{B}$, induced by a single infinitely long cylinder 
with radius $R$ 
and uniform magnetization, $\boldsymbol{M}=M\unitvector{x}$, perpendicular to its axis 
is~\cite{gerber1983,stratton1941} (Fig.~\ref{supp-fig:magnfieldforce_singlecylinder}a)
\begin{align}
\begin{split}
  \boldsymbol{B}_\text{single}(r,\theta) 
    &=\frac{\mu_0M}{2}\left(\frac{R}{r}\right)^2 
      \left[ \cos\theta\, \unitvector{r} + \sin\theta\, \unitvector{\theta} \right]
    \\&=\frac{\mu_0M}{2}\left(\frac{R}{r}\right)^2 
      \left[ \cos2\theta\,\unitvector{x} +\sin2\theta\,\unitvector{y} \right]
\,,
\label{supp-eq:magneticfield-final}
\end{split}
\end{align}
outside of the cylinder, where $\mu_0$ is the permeability of free space,
$r$ and $\theta$ are the plane-polar coordinates centered on the cylinder
and $\unitvector{x}$, $\unitvector{y}$ and $\unitvector{r}$, $\unitvector{\theta}$
are the unit vectors in Cartesian and plane-polar coordinates, respectively
(see~Fig.~\ref{supp-fig:coordinates-cylinder}).

We place into this field a spherical paramagnetic particle. 
If this particle is uniformly magnetized then it behaves precisely like a dipole of 
moment $\boldsymbol{m}$.
The force on a magnetic dipole $\boldsymbol m$ in a given magnetic field $\boldsymbol B$ 
is given by~\cite{griffithsdj1999}
\begin{align}
\boldsymbol{F}_\text{m}&=\nabla\left(\boldsymbol{m}\cdot\boldsymbol{B}\right).
\end{align}
Since the sphere is paramagnetic, its dipole moment is aligned with 
the surrounding 
magnetic field.
We assume that the magnetic field of the cylinder is just a small perturbation of a
stronger uniform outer field, aligned with the cylinder magnetization, \emph{i.e.},
\begin{align}
 \boldsymbol{m}&=m\unitvector{x}\,.
\end{align}
A uniform field does not create a force on the dipole,
thus, using the magnetic field \eqref{supp-eq:magneticfield-final}, 
we obtain (Fig.~\ref{supp-fig:magnfieldforce_singlecylinder}b)
\begin{align}
\begin{split}
 \boldsymbol{F}_\text{m, single}=\nabla\left(\boldsymbol{m}\cdot\boldsymbol{B}\right)
  &=\frac{\mu_0mMR^2}{2}\nabla\left(  \frac{\cos2\theta}{r^2} \right)
  =-\frac{\mu_0mMR^2}{r^3}\left( \cos2\theta\,\unitvector{r} + 
    \sin2\theta\,\unitvector{\theta}\right)
  \\&=-\frac{\mu_0mMR^2}{r^3}\left( \cos3\theta\,\unitvector{x} + 
    \sin3\theta\,\unitvector{y}\right)\,.
\end{split}
\end{align}

In setup~B, the magnetization of the cylinder
and the magnetic dipole moment of the particle are rotated by  $\pi/2$ and aligned with the $y$-axis,
\emph{i.e.}, $\boldsymbol{M}=M\unitvector{y}$ and $\boldsymbol{m}=m\unitvector{y}$.
In this case, the sign of the force swaps (compare Fig.~\ref{supp-fig:magnfieldforce_singlecylinder}b). 
Thus, we can express both setups in one formula as
\begin{align}
 \boldsymbol{F}_\text{m, single}
  &=\mp\frac{\mu_0 mM}{R}\left(\frac{R}{r}\right)^3
  \begin{pmatrix}
   \cos3\theta\\\sin3\theta
  \end{pmatrix}\,.
\end{align}


\section{Nondimensionalization and Mason number}
Neglecting inertial terms, 
the net force on the particle must be zero, 
\emph{i.e.}, the magnetic force, $\boldsymbol{F}_\text{m}$,
and the viscous drag force (Stokes drag) due to the flow field, $\boldsymbol{u}$,
must balance, giving an ODE for the particle trajectory, $\boldsymbol{x}_\text{p}(t)$,
\begin{align}
 6\pi\eta a \left( \boldsymbol{u}(\boldsymbol{x}_\text{p})
    -\frac{\partial\boldsymbol{x}_\text{p}}{\partial t}\right)
    +\boldsymbol{F}_\text{m}&=0,
\label{supp-eq:dimforcebalance}
\end{align}
with $\eta$ the  fluid viscosity
and  $a$ the particle radius.

Scaling the velocities with the maximum fluid velocity, $U$, 
and the magnetic force with $\mu_0 mM/R$,
\emph{i.e.}, the maximum magnetic force a single cylinder 
could exert on the particle,
and choosing a suitable length scale ${\L}$
for the geometry of the cylinder array
we get
\begin{align}
x &= {\L}\hat{x}
\,, & 
y &= {\L}\hat{y}
\,, & 
\boldsymbol{x}_\text{p} &= {\L}\hat{\boldsymbol{x}}_\text{p}
\,, & 
\boldsymbol{u} &= U\hat{\boldsymbol{u}}
\,, &
t &= \frac{{\L}}{U}\hat t
\,, &
\boldsymbol{F}_\text{m} &= \pm\frac{\mu_0 mM}{R}\hat{\boldsymbol{F}}_\text{m}
\,.
\end{align}
Thus the force balance can be rewritten in dimensionless terms as
\begin{align}
\frac{\partial {\hat{\boldsymbol{x}}}_p}{\partial \hat t}
  &= \hat{\boldsymbol{u}}
 +\frac{1}{\Mn}\hat{\boldsymbol{F}}_\text{m}
\end{align}
with the Mason number~\cite{kang2012}
\begin{align}
 \Mn&=\pm\frac{ \left| \boldsymbol{F}_\text{v} \right|}{ \left| \boldsymbol{F}_\text{m} 
    \right|}
  =\pm\frac{6\pi\eta a RU}{\mu_0 mM}.
\end{align}
The sign in the Mason number allows us to easily distinguish between setups A and B,
while using the same dimensionless magnetic force, $\hat{\boldsymbol{F}}_\text{m}$.

\section{Critical Mason number in infinite square cylinder array}

Due to symmetry, 
along a midline between two columns of cylinders,  \emph{e.g.}, at $\hat x=0$,
both the dimensionless magnetic force and the dimensionless fluid velocity 
only have components
in the $\hat y$-direction.
Thus, we can consider them as functions of $\hat y$ only,
say  $\hat F_{\hat y}(\hat y)$ and $\hat v(\hat y)$, respectively,
and a particle on the midline will in theory remain there for all time.
If this particle escapes,
its velocity,  
${\partial \hat y_\text{p}}/{\partial \hat t} = \hat v(\hat y)+\hat F_{\hat y}/\Mn$,
is always positive.
If it cannot escape, this means that 
${\partial \hat y_\text{p}}/{\partial \hat t}\le 0$ 
along part of the $\hat y$-axis.
At the critical Mason number, there just about exists a stationary point on the  
$\hat y$-axis,
that is 
\begin{align}
  &\exists\, \hat y^\star:\quad 0=\hat v(\hat y^\star)+\frac{1}{\Mn}
    \hat  F_{\hat y}(\hat y^\star) \,,
  &\text{while}\quad
  &0\le\hat v(\hat y)+\frac{1}{\Mn}\hat  F_{\hat y}(\hat y) 
    \quad \forall\,\hat y\,,
\intertext{or}
  &\exists\, \hat y^\star:\quad|\Mn|
    =\mp\frac{\hat  F_{\hat y}(\hat y^\star)}{\hat v(\hat y^\star)} \,
  &\text{and}\quad
    &|\Mn|\ge\mp\frac{\hat  F_{\hat y}(\hat y)}{\hat v(\hat y)} 
    \quad \forall\, \hat y\,.
\end{align}
Since $\hat  F_{\hat y}(\hat y)$  and $\hat v(\hat y)$ 
are odd and even functions respectively,
this absolute value of the Mason number 
is independent of the sign, 
and so does not depend on 
the setup.
Hence, we define the critical Mason number as
\begin{align}
 \Mnc\mathrel{\mathop:}=|\Mn|&=\max_{\hat y} 
    \left(\mp\frac{\hat  F_{\hat y}(\hat y)}{\hat v(\hat y)}\right).
\label{supp-eq:Mncritical}
\end{align}
If $|\Mn|\le\Mnc$ then we can guarantee full capture;
if $|\Mn|>\Mnc$ then some  particles will escape the cylinder array.

\section{Derivation of the asymptotic critical Mason number from lubrication theory}
The flow through two very close cylinders can be approximated
using lubrication theory~\cite{ockendon1995}.
With $R$ the cylinder radius and $2d$ the minimum cylinder gap thickness, 
the width of the gap along the $y$-axis is $2h$, with
\begin{align}
  h(y)&=d+R-\sqrt{R^2-y^2}
    \sim \left(1+\frac{y^2}{2Rd}\right)d
    +\mathcal{O}\left(y^3\right)
    \,.
\end{align}
In $x$-direction, we nondimensionalize with $d$.
Since the gap width, $2h$, varies on the scale $\sqrt{Rd}$, 
we use this scale to nondimensionalize $y$.
Upon nondimensionalizing Stokes equations with 
\begin{align}
x&=d\hat x \,,
 \qquad
 y={\L}\hat y \,,
&
u&=\sqrt{\frac{d}{R}} U\hat u  \,,
& 
v&= U \hat v \,,
\\
h(y)&\sim d\hat h(\hat y)=d\left(1+\frac{\hat y^2}{2}\right)  \,,
& 
p&=p_0+\frac{\eta U\sqrt{R}}{d^{3/2}}\hat p \,,
\end{align}
we obtain at leading order
\begin{align}
  0  &=-\frac{\partial \hat p}{\partial \hat x}
\,, &
  0  &=-\frac{\partial \hat p}{\partial \hat y}
    +\frac{\partial^2\hat v}{\partial \hat x^2}
\,, &
0&=\frac{\partial \hat u}{\partial \hat x}+\frac{\partial \hat v}{\partial \hat y}
\,,
\end{align}
with boundary conditions
\begin{align}
 \hat u&=0\,,\quad \hat v=0  &&\text{at}\quad \hat x=\hat h(\hat y)\,,&
 \hat u&=0\,,\quad\frac{\partial \hat v}{\partial \hat x}=0&&\text{at}\quad \hat x=0\,.
\end{align}
This can be solved by
integrating across the thickness
to obtain
\begin{align}
\label{supp-eq:ndflowvelocity}
 \hat u (\hat x,\hat y)&=\hat h^\prime
  \left[ \frac{\hat x}{\hat h^2} - \frac{\hat x^3}{\hat h^4} \right]
\,, &
 \hat v (\hat x,\hat y)&=\frac{1}{\hat h^3}\left[ \hat h^2-\hat x^2 \right]
\,, &
 \frac{d \hat p}{d \hat y}&= -\frac{2}{\hat h^3}
\,,
\end{align}
where $\hat h=\hat h(\hat y)$ and the prime denotes differentiation.

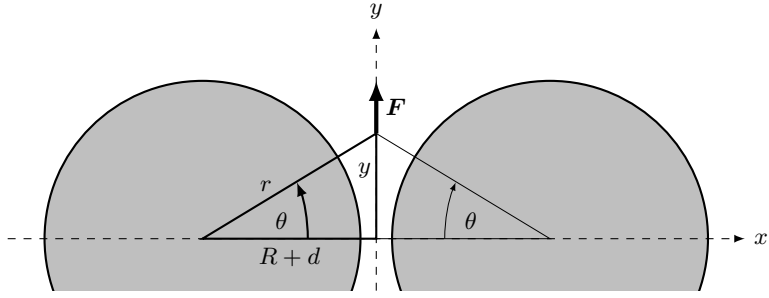
\begin{figure}\centering
  \begin{tikzpicture}[thick,x=0.7cm,y=0.7cm,>=latex]
    \newcommand{\R}{3}
    \renewcommand{\d}{0.3}
    \newcommand{\y}{2}
    \newcommand{\ymax}{4}
    \newcommand{\xmax}{7}

    \draw[fill=lightgray] (\R+\d,0) circle (\R);
    \draw[fill=lightgray] (-\R-\d,0) circle (\R);

    \draw[thin,dashed,->] (0,-\ymax) -- (0,\ymax);
    \node at (0,\ymax) [above] {$y$};
    \draw[thin,dashed,->] (-\xmax,0) -- (\xmax,0);
    \node at (\xmax,0) [right] {$x$};

    \draw[ultra thin]  (0,\y) -- (+\R+\d,0) -- (0,0);
    \draw[thick]       (0,\y) -- (-\R-\d,0) -- (0,0) -- (0,\y);
    \node at (-\R/2-\d/2,0) [below] {$R+d$};
    \node at (-\R/2-\d/2,\y/2) [left] {$r~$};
    \node at (0,\y*0.65) [left] {$y\!$};
    
    \node at  ( 0.5*\R + \d,0) [above] {$\theta$};
    \node at  (-0.5*\R - \d,0) [above] {$\theta$};
    \draw [ultra thin,->] ( \d+\R/3,0) arc (180:158:2cm);
    \draw [thick,->]      (-\d-\R/3,0) arc (  0: 22:2cm);

    \draw [ultra thick,->] (0,\y) -- (0,\y+1);
    \node at (0,\y+0.5) [right] {$\boldsymbol{F}$};

    \path[fill=white] (-\xmax,-\ymax) rectangle (\xmax+0.5,-1);
    \pgfresetboundingbox
    \path[use as bounding box] (-\xmax,-1) rectangle (\xmax+0.5,\R+0.3);
  \end{tikzpicture}
  \caption{Geometry  along the $y$-axis.}
  \label{supp-fig:geometrymidline}
\end{figure}
Following equation~\eqref{supp-eq:Mncritical}, it is only the midline that concerns us here,
\emph{i.e.}, (Fig.~\ref{supp-fig:geometrymidline})
\begin{align}
 x&=0\,,
&
y&= (R+d)\tan{\theta}\,,
&
r&=\frac{R+d}{\cos\theta}\,,
\intertext{thus}
\hat x&=0\,,
&
 \hat y&= \left(\sqrt{\frac{R}{d}}+\sqrt{\frac{d}{R}}\right)\tan\theta,
&
\hat r&=\frac{r}{R}=\frac{1+d/R}{\cos\theta}\,,
\end{align}
so that the vertical fluid velocity from  \eqref{supp-eq:ndflowvelocity} is 
\begin{align}
 \hat v(\theta)&=\frac{1}{\hat h}
  =\frac{1}{1+\frac{\hat y^2}{2}}
  = \left( 1+\frac{1}{2}  
    \left(\sqrt{\frac{R}{d}}+\sqrt{\frac{d}{R}}\right)^2\tan^2\theta \right)^{-1}
\end{align}
and, 
considering only these two cylinders, 
the $\hat y$-component of the magnetic force is
\begin{align}
 \hat{F}_{\hat y}(\theta)
  &= -\frac{2}{(1+d/R)^3}    \cos^3{\theta}\sin3\theta\,.
\end{align}
Let 
\begin{align}
g(\theta) &= - \frac{\hat{F}_{\hat y}(\theta)}{\hat v(\theta)}\,,
\end{align}
{then}
\begin{align}
\Mnc&=\max_{\theta\in(-\frac{\pi}{2},\frac{\pi}{2})}g(\theta)\,.
\end{align}
Hence
\begin{align}
\begin{split}
 g(\theta)&
  =\frac{2}{\left( 1+d/R \right)^3}\cos^3\theta\sin 3\theta
  \left( 1+\frac{1}{2}  
    \left(\sqrt{\frac{R}{d}}+\sqrt{\frac{d}{R}}\right)^2\tan^2\theta \right)
\\&=\frac{2}{\left( 1+d/R \right)^3}\cos^3\theta\sin 3\theta
  +\frac{1}{\left( 1+d/R \right)(d/R)}\cos\theta\sin^2\theta\sin 3\theta\,.
\end{split}
\end{align}
However, we can improve our understanding, by noting that for $d/R \rightarrow 0$,
${2/\left(1+d/R \right)^3\to 2}$, 
whereas 
${1/\left(\left( 1+d/R \right)(d/R)\right) \sim (d/R)^{-1}}$,
so that 
\begin{align}
 g(\theta)\sim \left(\frac{d}{R}\right)^{-1} \tilde g(\theta)=\left(\frac{d}{R}\right)^{-1}\cos\theta\sin^2\theta\sin 3\theta
\end{align}
and hence
\begin{align}
\begin{split}
 g^\prime(\theta)\sim \left(\frac{d}{R}\right)^{-1}  \tilde g^\prime(\theta) 
  &= \left(\frac{d}{R}\right)^{-1} \left[  
    -\sin^3\theta\sin3\theta
    +2\cos^2\theta\sin\theta\sin3\theta\right.\\&\qquad\qquad\left.
    +3\cos\theta\cos3\theta\sin^2\theta
  \right]
\\
  &=\left(\frac{d}{R}\right)^{-1}\sin^2\theta \left[
    2+4\cos2\theta+3\cos4\theta
  \right]
\\
  &=\left(\frac{d}{R}\right)^{-1}\sin^2\theta \left[
    24\cos^4\theta-16\cos^2\theta+1
  \right]\,.
\end{split}
\end{align}
A maximum of $g(\theta)$ has $g^\prime(\theta^\star)=0$ 
and so either $\theta^\star=0$ (not a max.)\
or 
\begin{align}
  24\cos^4\theta^\star-16\cos^2\theta^\star+1&=0\,.
\end{align}
Thus follows that for ${d/R}$ small enough, 
\begin{align}\begin{split}
 \Mnc &=
\max_{\theta}{g(\theta)}=
  g(\theta^\star)\sim\frac{1}{216}\left(34\sqrt{2}+5\sqrt{5}\right)\left(\frac{d}{R}\right)^{-1}
\approx
0.2744\left(\frac{d}{R}\right)^{-1}\,.
\end{split}
\end{align}

\end{bibunit}


\begin{thebibliography}{10}

\bibitem{yavuz2009}
Cafer~T. Yavuz, Arjun Prakash, J.~T. Mayo, and Vicki~L. Colvin.
\newblock Magnetic separations: From steel plants to biotechnology.
\newblock {\em Chem. Eng. Sci.}, 64:2510--2521, 2009.

\bibitem{ambashta2010}
Ritu~D. Ambashta and Mika Sillanp\"{a}\"{a}.
\newblock Water purification using magnetic assistance: A review.
\newblock {\em J. Hazard. Mater.}, 180(1--3):33--49, August 2010.

\bibitem{hayashi2011a}
S.~Hayashi, F.~Mishima, Y.~Akiyama, and S.~Nishijima.
\newblock Development of superconducting high gradient magnetic separation
  system for highly viscous fluid for practical use.
\newblock {\em Physica C}, 471(21–22):1511--1515, 2011.
\newblock The 23rd International Symposium on Superconductivity.

\bibitem{sinha2007}
Ashok Sinha, Ranjan Ganguly, Anindya~K. De, and Ishwar~K. Puri.
\newblock Single magnetic particle dynamics in a microchannel.
\newblock {\em Phys. Fluids}, 19(11):117102, 2007.

\bibitem{tsai2011a}
S.~S.~H. Tsai, I.~M. Griffiths, and H.~A. Stone.
\newblock Microfluidic immunomagnetic multi-target sorting -- a model for
  controlling deflection of paramagnetic beads.
\newblock {\em Lab Chip}, 11:2577--2582, 2011.

\bibitem{metso-spec-cyclic}
Metso Corporation.
\newblock {\em High gradient magnetic separators: HGMS cyclic}.
\newblock Technical specification: TS HGMS cyclic 1105-en.

\bibitem{pfister1979}
H.~Pfister.
\newblock {M}agnetische {S}eparation mit hohen {F}lussdichte-{G}radienten.
\newblock {\em J. Magn. Magn. Mater.}, 13(1-2):1--10, Sep 1979.

\bibitem{gerber1983}
Richard Gerber and Robert~R. Birss.
\newblock {\em High gradient magnetic separation}.
\newblock Electronic \& electrical engineering research studies. Magnetic
  materials and their applications series. Research Studies Press, 1983.

\bibitem{cummings1976}
Daniel~L. Cummings, David~A. Himmelblau, John~A. Oberteuffer, and Gary~J.
  Powers.
\newblock Capture of small paramagnetic particles by magnetic forces from low
  speed fluid flows.
\newblock {\em AIChE J.}, 22(3):569–575, 1976.

\bibitem{ebner2001}
Armin~D. Ebner and James~A. Ritter.
\newblock New correlation for the capture cross section in high-gradient
  magnetic separation.
\newblock {\em AIChE J.}, 47(2):303–313, 2001.

\bibitem{mishima2012}
F.~Mishima, S.~Hayashi, Y.~Akiyama, and S.~Nishijima.
\newblock Development of a superconducting high gradient magnetic separator for
  a highly viscous fluid.
\newblock {\em IEEE Trans. Appl. Superconduct.}, 22(3):3700204--3700204, June
  2012.

\bibitem{uchiyama1977a}
S.~Uchiyama, S.~Kurinobu, M.~Kumazawa, and M.~Takayasu.
\newblock Magnetic particle buildup processes in parallel stream type hgms
  filter.
\newblock {\em IEEE Trans. Magn.}, 13(5):1490--1492, Sep 1977.

\bibitem{chenf2012}
Fei Chen, Kenneth~A. Smith, and T.~Alan Hatton.
\newblock A dynamic buildup growth model for magnetic particle accumulation on
  single wires in high-gradient magnetic separation.
\newblock {\em AIChE J.}, 58(9):2865--2874, 2012.

\bibitem{lindner2013b}
Johannes Lindner, Katharina Menzel, and Hermann Nirschl.
\newblock Simulation of magnetic suspensions for {HGMS} using {CFD}, {FEM} and
  {DEM} modeling.
\newblock {\em Comput. Chem. Eng.}, 54:111 -- 121, 2013.

\bibitem{too1986}
C.~O. Too, M.~R. Parker, R.~Gerber, and D.~Fletcher.
\newblock Optimisation of matrix design in high gradient magnetic separation.
\newblock {\em J. Phys. D: Appl. Phys.}, 19(1):L1, 1986.

\bibitem{kimyg2013}
Y.~G. Kim, J.~B. Song, D.~G. Yang, J.~S. Lee, Y.~J. Park, D.~H. Kang, and H.~G. Lee.
\newblock Effects of filter shapes on the capture efficiency of a
  superconducting high-gradient magnetic separation system.
\newblock {\em Supercond. Sci. Technol.}, 26(8):085002, 2013.

\bibitem{hayashi1980}
K.~Hayashi and S.~Uchiyama.
\newblock On particle trajectory and capture efficiency around many wires.
\newblock {\em IEEE Trans. Magn.}, 16(5):827--829, 1980.

\bibitem{simons1980}
William~H. Simons and Richard~P. Treat.
\newblock Particle trajectories in a lattice of parallel magnetized fibers.
\newblock {\em J. Appl. Phys.}, 51(1):578, 1980.

\bibitem{mariani2010}
Giacomo Mariani, Massimo Fabbri, Francesco Negrini, and Pier~Luigi Ribani.
\newblock High-gradient magnetic separation of pollutant from wastewaters using
  permanent magnets.
\newblock {\em Sep. Purif. Technol.}, 72(2):147 -- 155, 2010.

\bibitem{yavuz2006}
C.~T. Yavuz, J.~T. Mayo, W.~W. Yu, and A.~Prakash.
\newblock Low-field magnetic separation of monodisperse {F}e$_3${O}$_4$
  nanocrystals.
\newblock {\em Science}, 314:964--967, 2006.

\bibitem{Note1}
See attached supplementary material for details of the
  computation of the critical Mason number and its asymptotic approximations.

\bibitem{kang2012}
Tae~Gon Kang, Martien~A. Hulsen, and Jaap M.~J. {den Toonder}.
\newblock Dynamics of magnetic chains in a shear flow under the influence of a
  uniform magnetic field.
\newblock {\em Phys. Fluids}, 24:042001, 2012.

\bibitem{freefemppdocumentation}
F.~Hecht.
\newblock {\em {F}ree{F}em++}, third edition, April 2014.
\newblock Version 3.30.

\bibitem{Note2}
Due to our choice of coordinate system, the ``capture distance'' or ``capture
  cross section'' from the literature\cite {cummings1976,simons1980} is
  $L-x_\protect \text {c}$.

\bibitem{ockendon1995}
H.~Ockendon and J.~R. Ockendon.
\newblock {\em Viscous flow}.
\newblock Cambridge University Press, 1995.

\end{thebibliography}

\begin{thebibliography}{1}

\bibitem{gerber1983}
Richard Gerber and Robert~R. Birss.
\newblock {\em High gradient magnetic separation}.
\newblock Electronic \& electrical engineering research studies. Magnetic
  materials and their applications series. Research Studies Press, 1983.

\bibitem{stratton1941}
Julius~Adams Stratton.
\newblock {\em Electromagnetic theory}.
\newblock McGraw--Hill Book Company, 1941.

\bibitem{griffithsdj1999}
David~Jeffrey Griffiths.
\newblock {\em Introduction to electrodynamics}.
\newblock Prentice Hall, third edition, 1999.

\bibitem{kang2012}
Tae~Gon Kang, Martien~A. Hulsen, and Jaap M.~J. {den Toonder}.
\newblock Dynamics of magnetic chains in a shear flow under the influence of a
  uniform magnetic field.
\newblock {\em Phys. Fluids}, 24:042001, 2012.

\bibitem{ockendon1995}
H.~Ockendon and J.~R. Ockendon.
\newblock {\em Viscous flow}.
\newblock Cambridge University Press, 1995.

\end{thebibliography}
\end{document}